\documentstyle[manuscript,aps]{revtex}
\begin{document}
\renewcommand{\theequation}{\thesection.\arabic{equation}}
\newcommand{\be}{\begin{equation}}
\newcommand{\ee}{\end{equation}}
\newcommand{\bea}{\begin{eqnarray}}
\newcommand{\eea}{\end{eqnarray}}

\title{Null Strings in Schwarzschild Spacetime}
\author{Mariusz P. D\c{a}browski \footnote{E-mail: mpdabfz@uoo.univ.szczecin.pl}\\
         {\it Institute of Physics, University of Szczecin, Wielkopolska 15,
          70-451 Szczecin, Poland}
{\ }\\
Arne L. Larsen \footnote{E-mail: alarsen@phys.ualberta.ca}\\
         {\it Department of Physics, University of Alberta, Faculty
         of Science, 412 Avadh Bhatia Physics Laboratory, Edmonton,
         Canada T6G 2J1}}
\date{\today}

\maketitle
\begin{abstract}
The null string equations of motion and constraints in the Schwarzschild
spacetime are given.
The solutions are those of the null geodesics of
General Relativity appended by a null string constraint in which the
"constants of motion" depend on the world-sheet spatial coordinate. 
Because of the extended nature of a string, the physical 
interpretation of the solutions is completely different from the point 
particle case. In particular, a null string is generally not propagating 
in a plane through the origin, although each of its individual points is.
Some special solutions are obtained and their physical interpretation is given. Especially, the solution for 
a null string with a constant radial coordinate $r$ moving vertically from 
the south pole to the north pole around the photon sphere, is presented. A general 
discussion of classical null/tensile strings as compared to massless/massive 
particles is given. For instance, tensile circular solutions with a constant 
radial coordinate $r$ do not exist at all. The results are discussed 
in relation to the previous literature on the subject.
\end{abstract}
\newpage

\section{Introduction}
\setcounter{equation}{0}
It is well known that the classical evolution of strings even in the simplest
curved backgrounds, such as the Schwarzschild spacetime, is described by a
complicated
system of second-order non-linear coupled partial differential equations. In
the Schwarzschild spacetime the system is actually non-integrable and it
subjects chaotic
behaviour \cite{arne}, so one may just try to find the exact evolution for some
special configurations \cite{fro,ini1,all,san1,san2,ini2} or perform some
numerical calculations \cite{arne,ini1,moss}.
This means that there
is no hope for making the full classification of the possible classical
trajectories of
strings in the Schwarzschild spacetime similar to the one for the point
particles; see for instance the standard textbook by Chandrasekhar 
\cite{chan}.

In the case of the null strings (tensionless strings) \cite{schild} this
situation is simplified
since the null strings, similarly to the massless point particles,
essentially sweep out the light
cone, and their equations of motion are essentially just geodesic equations of
General
Relativity appended by an additional constraint. General relativistic first
integrals for point particles  are known for most of the symmetric spacetimes,
and we can apply them
to null strings with almost no hesitation. Then, depending on the assumed shape
of a 
null string, in principle, one can solve the null string equations of motion in
many
cases. Such calculations have been performed recently, though not completely, 
by Kar \cite{kar} for Minkowski, Rindler, Schwarzschild and Robertson-Walker
spacetimes following an earlier idea originally suggested by Roshchupkin and 
Zheltukhin \cite{alex} for Robertson-Walker spacetimes.

The task of this paper is to discuss the null string evolution in the
Schwarzschild spacetime in more detail, and thereby also to shed light on the
solutions obtained by Kar \cite{kar}. We present the general equations of
motion for strings in the Schwarzschild spacetime and give the general 
solutions in quadratures in the case of null strings in Section II. In 
Section III
we solve the equations of motion completely in closed form for circular null
strings and we discuss their physical interpretation. In Section IV
we present a very interesting exact solution which describes a string moving
vertically up and down around the photon sphere.  In Section V we briefly
discuss the relation
between the tensile and the null strings in the context of our solutions.
Finally in Section VI, we summarize our results and give some concluding
remarks.

\section{Strings in the Schwarzschild spacetime}
\setcounter{equation}{0}
In any curved spacetime the spacetime coordinates describing a
string-configuration, in general, depend
on both of the string coordinates $\tau$ and $\sigma$, so we may use the
following
notation
\begin{equation}
X^{0} = t(\tau,\sigma), \hspace{0.5cm} X^{1} = r(\tau,\sigma),
\hspace{0.5cm} X^{2} = \theta(\tau,\sigma),
\hspace{0.5cm} X^{3} = \varphi(\tau,\sigma)  .
\end{equation}

Let us consider the tensile string (finite tension)  and the null string (zero
tension)
equations of motion in a compact formula:
\be
\ddot{X}^{\mu} + \Gamma^{\mu}_{\nu\rho} \dot{X}^{\nu}\dot{X}^{\rho}
= \lambda \left( {X}^{\prime\prime\mu} + \Gamma^{\mu}_{\nu\rho}
X^{\prime\nu} X^{\prime\rho} \right)   .
\ee
The constraints read as
\begin{eqnarray}
g_{\mu\nu}\dot{X}^{\mu}\dot{X}^{\nu} & = & - \lambda
g_{\mu\nu} X^{\prime\mu} X^{\prime\nu}   ,\\
g_{\mu\nu}\dot{X}^{\mu}X^{\prime\nu} & = & 0   .
\end{eqnarray}
Here $(...)^{\cdot} \equiv \partial/\partial \tau$ and  $(...)^{\prime} \equiv
\partial/\partial \sigma $.
For $\lambda = 1$ we have the tensile strings while $\lambda = 0$ applies for
the
null strings. In Refs. \cite{vega1,vega2,carlos,alex1}, expansion schemes were
considered, essentially using $\lambda$ as a continuous expansion parameter;
here we simply use $\lambda$ as a discrete parameter discriminating between
tensile and null strings.

>From the above we can see that for the null strings we have the
null geodesic equations supplemented
by the constraint (II.4), which ensures that each point of the string
propagates in the direction perpendicular to the string.  In the Schwarzschild
spacetime we have from (II.2)
\begin{eqnarray}
\ddot{t} - \lambda t^{\prime\prime} + 2 \frac{\frac{M}{r^2}}{1 - \frac{2M}{r}}
\left( \dot{t}\dot{r} - \lambda t^{\prime}r^{\prime} \right) & = & 0   ,\\
\ddot{r} - \lambda r^{\prime\prime} -  \frac{\frac{M}{r^2}}{1 - \frac{2M}{r}}
\left( \dot{r}^2 - \lambda r^{\prime 2} \right) +  \frac{M}{{r^2}}
\left(1 -
\frac{2M}{r}\right) \left( \dot{t}^2 - \lambda t^{\prime 2} \right)
 & - &  \nonumber \\
r \sin^2{\theta}\left(1 - \frac{2M}{r}\right) \left( \dot{\varphi}^2 -
\lambda \varphi^{\prime 2} \right) - r \left(1 - \frac{2M}{r}\right)
\left( \dot{\theta}^2 - \lambda \theta^{\prime 2} \right) & = & 0   ,\\
\ddot{\varphi} - \lambda \varphi^{\prime\prime} +  \frac{2}{r} \left( \dot{r}
\dot{\varphi} - \lambda r^{\prime} \varphi^{\prime} \right) +
2 \frac{\cos{\theta}}{\sin{\theta}} \left( \dot{\theta} \dot{\varphi} -
\lambda \theta^{\prime} \varphi^{\prime} \right) & = & 0   ,\\
\ddot{\theta} - \lambda \theta^{\prime\prime} +  \frac{2}{r}
\left( \dot{r}\dot{\theta} - \lambda r^{\prime} \theta^{\prime} \right) -
\sin{\theta} \cos{\theta} \left( \dot{\varphi}^2 - \lambda
\varphi^{\prime 2} \right) & = & 0  .
\end{eqnarray}
In the case of the null strings ($\lambda = 0$)
the equations (II.5) and (II.7) easily integrate. The only difference from that 
of the general relativistic point particle case is that now the "constants of
motion" must depend on the string coordinate $\sigma$, i.e.,
\begin{eqnarray}
\dot{t} & = & \frac{E(\sigma)}{1 - \frac{2M}{r}}   ,\\
\dot{\varphi} & = & \frac{L(\sigma)}{r^2 \sin^2{\theta}}    .
\end{eqnarray}
Combining (II.8) with (II.9) and (II.10) we obtain for $\lambda = 0$
\begin{equation}
r^4 \sin^2{\theta} \ddot{\theta} + 2 r^3 \dot{r} \sin^2{\theta} \dot{\theta}
- L^2(\sigma) \frac{\cos{\theta}}{\sin{\theta}} = 0    ,
\end{equation}
which integrates in a standard way
\begin{equation}
r^4 \sin^2{\theta} \dot{\theta}^2 = - L^2(\sigma) \cos^2{\theta} + K(\sigma)
\sin^2{\theta}     ,
\end{equation}
and the non-negative function $K(\sigma)$ generalizes Carter's `fourth constant' of motion (see for
instance \cite{mtw}).
The standard potential equation for the radial coordinate is then
obtained by integrating (II.6) in the case $\lambda = 0$
\begin{equation}
\dot{r}^2 + V(r) = 0     ,
\end{equation}
where
\begin{equation}
V(r) = -E^2(\sigma) + \frac{1}{r^2}\left(1 -
\frac{2M}{r}\right) \left[ L^2(\sigma) + K(\sigma) \right]   ,
\end{equation}
and
the constraint (II.3) has been taken into account. As we can easily see  
$L(\sigma)$ refers directly to the coordinate $\varphi$ while $K(\sigma)$ 
refers directly to the coordinate $\theta$.
One can also define the generalized impact parameter as \cite{chan}
\begin{equation}
D(\sigma) \equiv \frac{\sqrt{L^2(\sigma) + K(\sigma)}}{E(\sigma)}    .
\end{equation}

It should be noticed that if there was not the "constant of motion"
$K(\sigma)$ , the only solution of (II.12) with
$L \neq 0$ would be given by $\theta = const. = \pi/2$. As it is very 
well known in the case of point particles one can of course put $K = 0$ 
without loss of generality. However, because of the fact that the string is 
an extended 
object this is not the case here, and the explicit dependence on the string 
coordinate $\sigma$ must be given. In particular, the null string does 
{\it not} in general move on a
plane through the origin, although each individual point  of the string
actually does.

The constraint (II.4) now takes the form
\begin{equation}
E(\sigma)t^{\prime} - \frac{\dot{r}}{1 - \frac{2M}{r}}r^{\prime} - L(\sigma)
\varphi^{\prime} - r^2 \dot{\theta} \theta^{\prime} \sin^2{\theta}  = 0   ,
\end{equation}
with $\dot{r}$ and $\dot{\theta}$ given by (II.12)-(II.13). Equation (II.16)
means that we have a constraint on the functions $E(\sigma)$,
$L(\sigma)$ and $K(\sigma)$.

Finally, we notice that the invariant string size (the length of the string) $S(\tau)$ for the null string
is given by
\begin{equation}
S(\tau)=\int_0^{2\pi}\;S(\tau,\sigma)d\sigma,
\end{equation}
where
\begin{equation}
 S(\tau,\sigma)= \sqrt{g_{\mu \nu}X^{\prime \nu}X^{\prime\nu}} = \left[-\left(1
-
\frac{1}{2M}\right)t^{\prime 2} + \left(1 -\frac{1}{2M}\right)^{-1}
r^{\prime 2} + r^2 \theta^{\prime 2} + r^2 \sin^2{\theta} \varphi^{\prime 2}
\right]^{1/2}   .
\end{equation}

\section{Circular null strings}
\setcounter{equation}{0}
As a first example \footnote{It is instructive to learn that a planetoid string 
ansatz applied in \cite{ini2} is not suitable for the null strings at all.} 
of exact solutions, we consider the circular ansatz for a null string 
in the Schwarzschild spacetime:
\begin{equation}
t = t(\tau) , \hspace{0.3cm} r = r(\tau), \hspace{0.3cm} \theta = \theta(\tau),
\hspace{0.3cm} \varphi = \sigma   .
\end{equation}

Inserting (III.1) into (II.9)-(II.13) we have
\bea
\dot{t} & = & \frac{E(\sigma)}{1 - \frac{2M}{r}}   ,\\
\dot{r}^2 & = & E^2(\sigma) - \frac{K(\sigma)}{r^2} \left( 1 - \frac{2M}{r}
\right)   ,\\
r^4 \dot{\theta}^2 & = & K(\sigma)   ,
\eea
and we have also used the constraint (II.16) which gives the condition for
$L$ to be equal to zero. It is also clear from (III.2)-(III.4) that
$E$ and $K$ must be {\it constants} (independent of $\sigma$) in this case.
This is not the case in general, of course. One can easily learn about it by 
taking $K = L = 0$ and $\theta = \pi/2$ to integrate (II.9), (II.10) and 
(II.13) completely having $E$ and some new 'constants of integration' (like 
$r_{0}$ and $t_{0}$ in (III.6) below) to be $\sigma-$dependent. 

The simplest solution of the set (III.2)-(III.4) come if we put $K = 0$, 
i.e.,
\bea
\theta & = & const.   ,\\
r - r_{0} + 2M \ln{ \frac{r - 2M}{r_{0} - 2M}} & = &
\pm \left( t - t_{0} \right)   ,
\eea
and they describe the "cone strings" which start with finite size and then
sweep out the cones $\theta=$ const. These strings play similar role as the 
radial null geodesics in General Relativity.
Notice that the extended nature of the string means that configurations
corresponding to different values of the constant polar angle $\theta$ are
physically different: for $\theta=\pi/2,$
the string is in a plane through the origin while for any other value of
$\theta$ it is not. In fact, for $\theta=\pi/2,$ the string winds around the
black hole in the equatorial plane while for $\theta\neq\pi/2,$ it is on a
parallel plane moving perpendicular to the equatorial plane. In contrast, a
point particle is always in a plane through the origin. This illustrates the 
fact that although the null string equations are similar to the massless 
geodesic equations, in particular when $E$ and $K$ are constants, the physical interpretation of the solutions is 
completely different.

For the sake of comparison we mention that in Minkowski spacetime $(M = 0)$ the
logarithmic term vanishes giving simply
\bea
\theta & = & const.   ,\\
r - r_{0} & = & \pm \left( t - t_{0} \right)   .
\eea
The "cone strings" also appear in the anti-de-Sitter spacetime since there
\bea
\theta & = & const.   ,\\
r - r_{0} & = & \pm \frac{1}{H} \tan{H \left( t - t_{0} \right)}   ,
\eea
with $H =$ const.$ = \sqrt{- \Lambda/3}$, where $\Lambda$ is the cosmological
constant. Such "cone strings" can also easily be constructed in other static
spherically symmetric spacetimes, but we shall not go into further details
here.

Coming back to the equations (III.2)-(III-4) for arbitrary $K,$ we first notice that (III.3) is 
exactly equivalent to the equation for photons moving in the equatorial plane 
with non-vanishing angular momentum $(L\neq 0,\;\theta=\pi/2\;\Rightarrow\;
K=0).$ This, of course, just reflects the fact that point particles always 
move in a plane through the origin. However, since a string (even a null 
string) is an extended object, the physical interpretation of the solutions 
to (III.3) is completely different from that of point particle solutions. 
In particular, none of the string solutions we will obtain here are 
propagating in a plane through the origin. But the qualitative and 
quantitative picture for the string solutions can still be extracted from 
the well-known results for point particles (see for instance \cite{chan,mtw}).
It is possible because in the assumed ansatz (III.1), all the spacetime 
coordinates 
are just functions of one of the two string coordinates, and we can derive 
the equation which relates the coordinates $r$ and $\theta$ from 
(III.2)-(III.3) in a similar way as one
usually does for the massless particles \cite{chan}, although now, instead of 
$\varphi$ we have $\theta$, and instead of $L$ we have $\sqrt{K}$, i.e.,
\bea
\left( \frac{du}{d\theta} \right)^2 & = & 2M\left(u + \frac{1}{6M}\right)
\left(u - \frac{1}{3M}\right)^2  
+\frac{1}{M^2}\left(\frac{M^2}{D^2}-
\frac{1}{27}\right)  \nonumber \\
& = & 2Mu^3 - u^2 + \frac{1}{D^2}   ,
\eea
where $u = 1/r$ and $D$ is defined by (II.15). From this equation we can 
immediately derive the trajectories of the null strings similarly as in 
\cite{chan}. In our discussion we just use directly the Eq.(III.3)
looking for the turning points ($\dot{r}=0)$ of it, fulfilling
\begin{equation}
\frac{K}{4M^2 E^2}=\frac{(r/2M)^2}{1-(r/2M)^{-1}}.
\end{equation}
The equation (III.12) has two solutions outside the horizon provided
\begin{equation}
\frac{K}{M^2E^2} = \frac{D^2}{M^2} > 27,
\end{equation}
one solution in case of equality and otherwise no solutions.
Consider first the case where $K<27 M^2E^2\;(D < 3\sqrt{3}M)$ and a circular 
string incoming from spatial infinity (say) $\theta=0,\,r=\infty.$  The plane 
of the string is always parallel to the equatorial plane and the string 
approaches the south pole of the black hole, and since there is no turning 
point in this case, the string will eventually fall into the black hole. If 
$K>27 M^2E^2\;(D > 3\sqrt{3}M),$ the string again approaches the south pole of 
the black hole, but in this case it will scatter off and escape towards 
infinity again. Notice that in both cases the string can make a number of 
turns, moving vertically from the south pole to the north pole and back again 
and so on, around the black hole (during which it actually collapses 
($r\sin\theta=0)$ several times), but always with its plane parallel to the 
equatorial plane, before its fate is determined. Similarly one can consider 
strings starting very close to (but outside) the horizon with increasing $r(\tau).$ If 
$K<27 M^2E^2,$ the string escapes to infinity while if $K>27 M^2E^2,$ it will 
hit the barrier and fall back into the black hole. In the next section
we will consider a limiting case of this kind of dynamics.

\section{Null strings on the photon sphere}
\setcounter{equation}{0}

In the special case when the impact parameter $D = 3\sqrt{3} M,$ the Eq.(III.11) 
(with the circular ansatz (III.1) valid) factorizes and the simplest solution 
for the constant radial coordinate $r = 1/u = 3M$ comes immediately. 
If $r = 3M$, then we conclude from (III.11) and (III.2) that 
\begin{equation}
t(\tau) = 3E\tau ,  \hspace{0.5cm}  D^2=27M^2   .
\end{equation}
Then, one is able to integrate (III.4) to give
\begin{equation}
\theta = \pm \frac{E\tau}{\sqrt{3}M} + \theta_{0}   ,
\end{equation}
and $\theta_{0} =$ const.
Finally, we conclude that a circular string solution is described explicitly by 
\begin{equation}
t = 3E\tau , \hspace{0.3cm} r = const. = 3M,
\hspace{0.3cm} \theta = \pm \frac{E\tau}{\sqrt{3}M} + \theta_{0}   ,
\hspace{0.3cm} \varphi = \sigma ,
\end{equation}
which means that the string may move vertically from the south pole to the
north pole and back again and so on around the photon 
sphere ($r=3M).$ Notice that the factor $3E$ in equation (IV.1) can be scaled away. The invariant
string size (II.17) of the string solution (IV.3) is simply
\begin{equation}
S = 6\pi M \sin{(\pm  \frac{E\tau}{\sqrt{3}M} + \theta_{0} )}   .
\end{equation}
This string solution with constant radial coordinate $r=3M$ is, besides the solution 
\begin{equation}
t=\tau,\;\;r=const.=2M,\;\;\theta=const.,\;\;\varphi=\sigma,
\end{equation}
considered by Kar \cite{kar}, the only string solution with constant $r.$ Notice that the solution $r=2M$  is stationary while the solution $r=3M$ is highly dynamical.
It must be stressed, however, that the solution $r=3M$ is unstable. The situation 
is similar to the case of a photon on a circular orbit at $r=3M$ 
\cite{chan,mtw} and it means that there exist two asymptotic solutions in 
addition to the one given by (IV.3), one of which describes an incoming string 
approaching the photon sphere from infinity ($r = \infty$), then spiralling 
around it infinitely many times, and the other one, an outgoing string starting from somewhere close to (but outside) the horizon approaching the photon sphere by also spiralling 
around it infinitely many times. The exact expressions for such trajectories have been already given in 
\cite{chan}. For the string approaching the photon sphere from infinity 
(the orbit of 'the first kind' as called in \cite{chan}) we have 
from (III.2)-(III.4)    
\bea
\frac{1}{r} & = & - \frac{1}{6M} + \frac{1}{2M} {\rm tanh}^2\left(\frac{1}{2} 
(\theta + \theta_{0}) \right)   ,\\
\frac{dt}{d\tau} & = & \frac{E}{\frac{4}{3} - {\rm tanh}^2\left(\frac{1}{2} 
(\theta + \theta_{0}) \right)}   ,\\
\frac{d\theta}{d\tau} & = & \pm \frac{E}{4\sqrt{3}M} 
\left[ - 1 + 3 {\rm tanh}^2 \left(\frac{1}{2} (\theta + \theta_{0}) \right) 
\right]^2   ,
\eea
and for the outgoing string approaching the photon sphere (the orbit of 
'the second kind') 
\bea 
\frac{1}{r} & = & \frac{1}{3M} + \frac{2z}{M(z - 1)^2}   ,\\
\frac{dt}{d\tau} & = & \frac{E}{\frac{1}{3} - {\rm tan}^2 
\left( 2{\rm arctan}(\sqrt{z}) \right)}   ,\\
\frac{dz}{d\tau}  & = & 
\pm \frac{E}{\sqrt{3} M}\;\frac{z\left[ z^2 + 4z + 1 \right]^2}
{(z - 1)^4} ,
\eea
where $z = \exp{\theta}$ and $\theta_{0} =$ const. The exact solutions 
of these equations are given in terms of elliptic functions and we will not 
discuss them further on since their physical meaning is clear from the 
qualitative considerations given above and at the end of Section III. 
One can easily see from (IV.6)-(IV.8) that in the limit $\theta \rightarrow \infty$, $r = 3M$, 
$t = 3E\tau$ and $d\theta/d\tau=\pm E/\sqrt{3}M.$ Similarly in (IV.9)-(IV.11) for  $\theta \rightarrow \infty$, $r = 3M$, 
$t = 3E\tau$ and $d\theta/d\tau=\pm E/\sqrt{3}M,$ thus in both cases we obtain the limiting case (IV.3) as we should.
 
\section{Tensile circular strings in the Schwarzschild background}
\setcounter{equation}{0}

In this Section we briefly consider the case of tensile strings and start with 
the circular ansatz of Section III given by Eq. (III.1). The equations of 
motion (II.5)-(II.8) are given by
\bea
\dot{t} - \frac{E}{1 - \frac{2M}{r}} & = & 0  ,\\
\ddot{r}  -  \frac{\frac{M}{r^2}}{1 - \frac{2M}{r}} \dot{r}^2  +
\frac{M}{{r^2}}\left(1 - \frac{2M}{r}\right) \dot{t}^2 +
r \sin^2{\theta}\left(1 - \frac{2M}{r}\right)  - r \left(1 -
\frac{2M}{r}\right) \dot{\theta}^2  & = & 0   ,\\
\ddot{\theta}  + 2 \frac{\dot{r}}{r} \dot{\theta} + \sin{\theta}
\cos{\theta} & = & 0   ,
\eea
and the constraint (II.4) is automatically fulfilled, while the constraint
(II.3) gives
\be
\frac{E^2 - \dot{r}^2}{\left( 1 - \frac{2M}{r} \right)} - r^2 \dot{\theta}^2 -
r^2 \sin^2{\theta} = 0    .
\ee
These equations can be easily integrated in the equatorial plane and the
solutions have been discussed in Refs.\cite{ini1,all,san1} so we shall not
repeat them here.
The first special solutions of (V.1)-(V.3) outside the equatorial plane 
one might look for, are the ones with
constant radial coordinate $r =$ const. (c.f. the discussion of the null
strings in the previous section). However, as one can easily check
by simple substitution $\dot{r} = 0$ in (V.1)-(V.3) and then differentiation of
(V.2) with respect to $\tau$, this results in contradiction with (V.3).
This means that tensile strings with a constant value of the radial coordinate
$r$ do not exist at all. This is a very big difference from the point particles 
since for massive particles there are circular orbits which are stable for 
$r > 6M$ and unstable for $r < 6M$. It seems that it is impossible to keep 
such a symmetric and stationary configuration because of the selfinteraction 
of the tensile strings.

A second first integral (besides (V.4)) of the system (V.1)-(V.3), which would
guarantee its full integrability, is not known. In fact, numerical
investigations strongly suggest that no such integral exists, i.e., the system
is chaotic. The numerical investigations showed that essentially three possible
evolution schemes are possible for the axisymmetric tensile string in the
Schwarzschild spacetime \cite{arne,ini1}: (1) the string simply passes the
horizon and falls onto the black hole. (2) the string passes by the black hole,
but part of its translational energy is transformed into oscillatory energy.
(3) the string is "trapped" jumping chaotically around the black hole for a
certain amount of time before it either falls into the black hole or escapes
towards infinity. Notice that this is qualitatively the same kinds of dynamics that 
we found for the null strings using exact analytical methods of 
Sections III and IV. This suggests that by  finetuning of the initial conditions,
it should actually be possible to have (tensile) string solutions jumping around the
black hole forever\footnote{We thank D. Page for interesting discussions on
this point}, somewhat similar to the null string on the photon sphere as
discussed in section IV, but with the radial coordinate not exactly constant.
In the present case of tensile strings it is however not known whether such
solutions would be stable under small perturbations, and their existence might
demand {\it infinite} finetuning of the initial conditions.
We finally notice that in the case of electrically charged strings, such
solutions actually do exist \cite{arne}; their stability being
guaranteed by a Coulomb barrier for small radius and a tension barrier for
large radius. For the ordinary uncharged tensile strings, such solutions have
not yet been found, neither by analytical nor by numerical methods.

\section{Summary}

After our analysis of the evolution of strings in the above cases we now come
to the following discussion which compares the behaviour of classical null
strings and tensile strings to the behaviour of
massless and massive point particles in curved spacetimes.

A classical massive point particle in a curved spacetime experiences only the
interaction
with the gravitational field. It means that taking the limit of zero mass
changes the worldline of the particle smoothly since photons subjects the same
rule and they also interact with the gravitational field only.
Thus from the point of view of worldlines only, the limit of zero mass is
relatively mild.
For strings the situation is somewhat more complicated. What distinguishes the
strings from the point particles in such
situations is that the strings not only interact with the gravitational field,
but also selfinteract due to the tension. This makes an important physical
difference with what we have for the point particles in the sense that now
taking the limit of zero tension is quite {\em dramatic}. It is because that in
this limit
the selfinteraction of strings totally vanishes, while the interaction with the
gravitational field changes smoothly (as for point particles).

It is thus obvious that the null-particles (like photons) have something to do
with
massive particles. The general question is then, whether the null strings
actually
have anything to do with the tensile strings. Referring to that,
there have been suggestions recently \cite{vega1,vega2,carlos,alex1} that the
tensile string
equations of motion may be expanded perturbatively in inverse string tension
parameter $\alpha^{\prime}$ (or a parameter related to it by rescaling) having
the null string equations of motion as the
zeroth order approximation. However, although the equations of the null strings
are mathematically much simpler than the equations of the tensile strings, one
can question (c.f. the above discussion)
whether it is physically meaningful to consider a null string as a zeroth
order approximation of a tensile string. It seems also that there is some
disagreement in the literature \cite{vega1,vega2,carlos,alex1} about how to
define such an expansion scheme correctly.

On the other hand, it is certainly meaningful and interesting to consider null
strings by themselves (assuming that such objects actually exist).  The
dynamics of null strings in a curved spacetime is however very simply obtained
if the dynamics of point particles is known. As discussed in this paper, it is essentially a question of interpreting the well-known point particle results in the framework of an extended object. 
Referring to our calculations one can distinguish two different physical
situations. These are when the constraint (II.4)  is or is not automatically
fulfilled by the ansatz. If it
is automatically fulfilled the evolution of the string is almost trivial, in
the sense that each point
of the string follows a null geodesic (a trajectory of a massless particle)
without any correlations with the rest of the string. Then, the string motion
reduces to the motion of a collection of massless point particles moving quite
{\em independently}. On the other hand, if the constraint (II.4) is not
automatically fulfilled by the ansatz then there are some non-trivial
correlations between
the different points of the string, the nature of which are purely 'stringy'.
It appears that the 'stringy' nature is absent in any axially symmetric
spacetime for the circular ansatz, thus to look
for more complicated dynamics
one should consider either some less symmetric  background spacetimes
or some different string shapes.

Similar problems as discussed here might appear for the null p-branes in curved
spacetimes
\cite{alex2} which generalizes strings to higher-dimensional objects.

\section{Acknowledgments}
The authors wish to thank Don Page and Alex Zheltukhin for useful discussions.
MPD is supported by the Polish Research Committee (KBN) grant No 2 PO3B 196 10.
ALL is supported by NSERC (National Sciences and Engineering Research 
Council of Canada).

\end{document}